# Photophysics of single nitrogen-vacancy centers in nanodiamonds coupled to photonic crystal cavities

Philip P. J. Schrinner†, Jan Olthaus‖, Doris E. Reiter‖, Carsten Schuck†*

†Institute of Physics & CeNTech-Center for Nanotechnology & SoN-Center for Soft Nanoscience, University of Münster,
48149 Münster, Germany
*e-mail address: carsten.schuck@uni-muenster.de
‖Institute of Solid State Theory, University of Münster
48149 Münster, Germany

The nitrogen vacancy center in diamond in its negative charge state is a promising candidate for quantum optic experiments that require single photon emitters. Important benefits of the NV center are its high brightness and photo-stability, even at room temperature. Engineering the emission properties of NV centers with optical resonators is a widely followed approach to meet the requirements for quantum technological applications, but the effect on non-radiative decay paths is yet to be understood. Here we report on modifying the internal quantum efficiency of a single NV center in a nanodiamond coupled to a 1D photonic crystal cavity. We assess the Purcell enhancement via three independent measurement techniques and perform autocorrelation measurements at elevated excitation powers in confocal microscopy. Employing a three-level model allows us to extract the setup efficiency, individual transition rates and thus the internal quantum efficiency of our system. Combining our results, we find that the enhancement of the radiative decay rate via the Purcell effect results in an internal quantum efficiency of 90 % for cavity-coupled NV centers. Our findings will facilitate the realization of nano-scale single photon sources with near-unity internal quantum efficiencies operating at high repetition rates.

## I. INTRODUCTION

Color centers in diamond are excellent single photon sources for quantum optic experiments, quantum sensing and quantum information processing devices[1,2]. Especially, the negatively charged nitrogen vacancy (NV-) center in diamond has enabled pioneering experiments[3,4] due to its stable optical emission. However, larger scale implementations of quantum optic (interference) experiments require higher emission rates and efficiencies. This has motivated efforts for engineering the emission properties of NV centers by utilizing the Purcell effect[5–7].

The emission rate of an emitter can be enhanced by coupling to an optical resonator, as described by the Purcell factor, which increases with decreasing mode volume of the cavity. Strong optical mode confinement can be achieved both with plasmonic and photonic structures. Whereas plasmonic structures achieve stronger confinement, they generally suffer from larger losses[8], which quickly compromise the quantum efficiency[9]. Here the (internal) quantum efficiency refers to the probability of the excited state to decay along a preferred radiative transition, instead of other, e.g. non-radiative, decay paths. Photonic crystal (PhC) cavities fabricated from dielectric thin films[10,11] on the other hand, can reach mode volumes below $V_{\text{mod}} \lesssim 10(\lambda/n)^3$ with negligible loss and correspondingly high quality factors ($Q > 10^3$).

To efficiently couple NV centers in diamond to such dielectric nanostructures, the emitter size should be smaller than the mode volume. This can be fulfilled when resorting to nanodiamonds, where however it was found that the optical properties, in particular the quantum efficiency, of embedded NV centers degrade with decreasing distance to the surface of the nanodiamond[12]. The degradation is here especially relevant for high excitation rates. This effect can be modeled by an additional, non-radiative third level (or shelving state) and was verified by second order auto-correlation measurements at elevated excitation powers[12–14]. However, these measurements crucially require low loss and low fluorescent slab and substrate materials in order to not obscure the single photon signal from the NV center. Customary material systems that satisfy these requirements, such as fused silica, unfortunately do not lend themselves straightforwardly to realizing tailored optical environments with nanophotonic functionality, such as photonic crystals, due to their relatively low refractive index.

Even though Purcell enhancement for cavity-coupled NV centers has been shown with various implementations and the three-level model description for NV centers in diamond nanocrystals is widely accepted, the interplay of the internal dynamics of a nanodiamond-embedded NV center with a coupled optical resonator mode has remained unknown.

Here we report on the Purcell effect on the internal quantum efficiency of a single NV center in a nanodiamond that is coupled to a PhC cavity, which is driven at elevated excitation powers.

We realize optical resonators with small mode volumes and high quality factors by fabricating 1D PhC cavities in tantalum pentoxide ($Ta_2O_5$) thin films on insulator, as shown schematically in fig. 1 a). Using a lithographical positioning technique, we place nanodiamonds with single NV centers at the center of a PhC cavity for efficient optical coupling, as shown in fig. 1 b). The $Ta_2O_5$ material choice satisfies the requirement for low self-fluorescence even under elevated levels of optical excitation, as shown in fig. 1 c), while also realizing high refractive index contrast with respect to a buried oxide layer for strong optical mode confinement.[15]

We characterize the system in confocal microscopy and extract a total Purcell enhancement factor of $F_{tot} \approx 3.5$ by employing three independent measurement techniques for confirming resonator coupling beyond the vacuum contribution ($F_{tot,vac} = 1$).

We further extract the setup efficiency and the individual rates of the three-level system from autocorrelation measurements at varying excitation power, which in turn allows for estimating the internal quantum efficiency of the cavity-coupled NV center.

Finally, we combine our findings and derive the Purcell effect on the internal quantum efficiency. We find that the non-radiative decay rate is unaffected by the Purcell effect, such that the enhancement of the radiative decay rate results in an overall increase in the internal quantum efficiency from 75% to 90%.

Our work thus shows that, in contrast to plasmonic systems[9,16], the non-radiative decay path remains unaffected by the cavity because of the low loss dielectric environment. Hence, nanodiamond-specific surface effects on the internal dynamics of an NV center can be compensated efficiently with a PhC cavity.

## II. SIMULATION

In order to enhance the emission rate of a single NV center by the Purcell effect, we investigate nanodiamonds positioned at the center of 1D PhC cavities. The latter are realized via interference from periodically repeating air holes patterned into a dielectric waveguide, thus acting as Bragg mirrors.

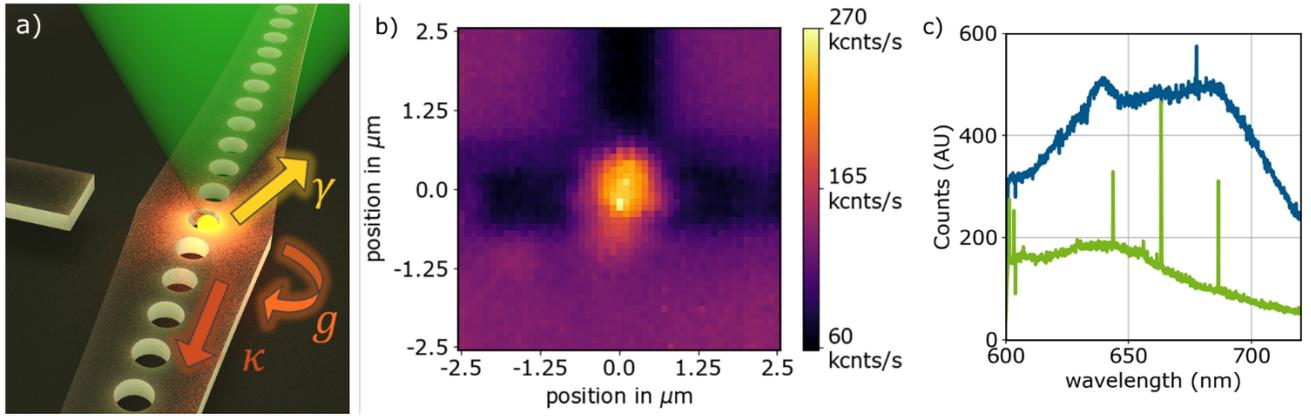

Figure 1: Setup and reference measurements performed in confocal microscopy. a) Schematic representation of a device consisting of a PhC cavity with embedded NV-center and perpendicular waveguide for optical excitation in cross-bar geometry. The NV center in a nanodiamond with decay rate $\gamma$ is coupled with a coupling strength $g$ to a 1D PhC cavity with decay rate $\kappa$. The NV center is excited with a green laser from top in the case of confocal microscopy measurements and radiates into the cavity. b) Photoluminescence map of a representative device (rotated by -90° with respect to a)) with 30 mW excitation power in confocal microscopy. The bright spot at the center is the NV- emitter and dark areas are the waveguides made of $Ta_2O_5$ c) Spectral distribution of self-fluorescence from $Ta_2O_5$ thin-films (green) and a single NV- emitter on $SiO_2$ (blue) under excitation with a 532 nm wavelength laser. The largest difference between the emission from NV- and $Ta_2O_5$ is observed from 650 nm to 700 nm wavelength.

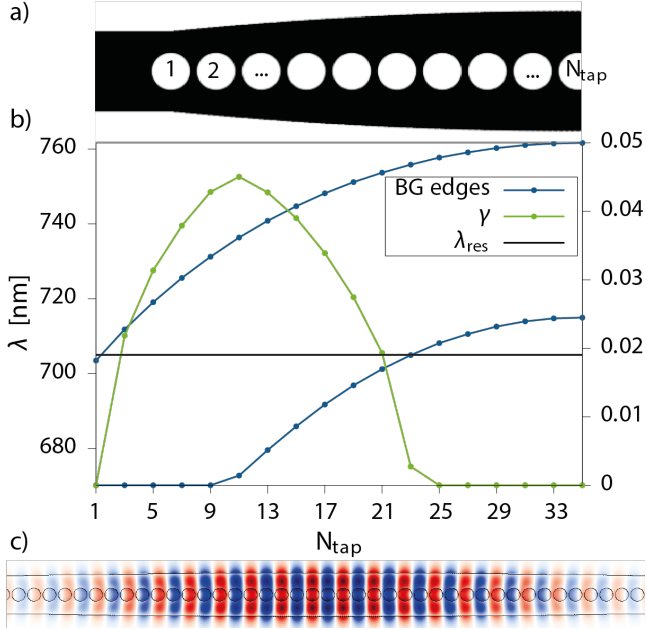

Figure 2: Bandgap position throughout the segments of the taper section. a) Schematic structure and definition of segment number. The defect center is located at $N_{tap} = 35$, the feeding dielectric $Ta_2O_5$ waveguide is located starting from $N_{tap} = 1$. b) Bandgap (BG) edges and mirror strength $\gamma$ throughout the cavity segments. Solid line shows resonant wavelength ($\lambda_{res}$ =705 nm) for reference. c) Electric field profile of the resulting defect mode at 705nm in the XY-plane.

Previously, attention has focused primarily on free-standing PhC cavities at telecom wavelengths for applications in optical communication and sensing[17]. Here instead we employ on-substrate PhC cavities[11] in 330 nm thick $Ta_2O_5$ waveguides on insulator for a target wavelength in the 650-750 nm range, which avoids spectral regions where $Ta_2O_5$ shows elevated levels of self-fluorescence under optical excitation, while taking advantage of the strong emission from NV centers at room temperature (see fig. 1 c).

Optimal coupling of a NV center to a $Ta_2O_5$ waveguide requires maximal spatial and spectral overlap of the emitter dipole with the resonant cavity mode. Here we achieve high spatial overlap using a deterministic air-mode PhC design[18] with hole diameters larger than 140 nm for reproducible fabrication. The design supports a bandgap for transverse electric (TE) modes, despite the relatively large thickness of the waveguide layer and the presence of the $SiO_2$ substrate. For optimal spectral overlap, the cavity linewidth has to fall within the emission spectrum of the NV center, which is easily satisfied at room temperature, where several phonon side bands broaden the emission spectrum. With both spatial and spectral conditions fulfilled, the Purcell factor then mainly depends on the quality factor to mode volume $Q/V_{mod}$ ratio of the cavity.

We perform band structure calculations using the frequency-eigensolver MPB in order to optimize the Purcell effect of our cavity designs. Using constant air hole distances ($a$ =230 nm) and radii ($r$ =95 nm) throughout the whole structure, a quadratic modulation of the waveguide width creates a defect. At the center of this defect the width reaches its maximum value ($w_{cen} = 644$ nm) and is tapered down to $w_{mir} = 430$ nm on both sides (see fig. 2 a). The TE-like bandgap in the taper section of our design is shown in fig. 2 b), which also shows the mirror strength (reflectivity)[19] of each segment.

Notably, we here optimize the design parameters such that the bandgap vanishes both in the vicinity of the defect center, and when approaching the feed waveguide. This allows for tuning the resonance wavelength (defect center) and minimizing losses at the PhC cavity-waveguide interface[20,21]. The resulting design shows a maximum TE-bandgap around 672-736 nm at the $N_{tap} = 11$ segment. Moreover, PhC parameters are chosen such that the resulting maximum bandgap is centered around our target frequency (maximizing mirror strength (reflectivity)), as discussed in further detail elsewhere.[11]

We then simulate the overall cavity geometry, rather than individual segments, for analyzing the coupling conditions. We employ 3D finite difference time domain (FDTD) simulations using the MIT Electromagnetic Equation Propagation package (MEEP).[22] To guarantee high transmission through the device, we set the total segment number to $N_{tap,max} = 35$. After parameter optimization we find a PhC cavity structure that supports a resonant mode at $\lambda_{res} = 705$ nm with Q=19800 and $V_{mod} = 6.25(\lambda/n)^3$.

Lastly, we analyze the perturbation of the optical mode in the PhC cavity arising from integration of nanodiamonds into the central air hole at the defect center of the cavity. We here consider diamond spheres of $r = 15 - 35$ nm radius and vary their position within the mode profile of the cavity.

For diamonds positioned (1) at the center of the central air hole, $Q/V_{mod}$ decreases slightly with increasing sphere radius, as shown for $r = 15, 25, 35$ nm in fig. 3. Interestingly, placement (2) at the edge of the central air hole results in a relatively large increase of the $Q/V_{mod}$ ratio with the sphere radius. At this position, the relative E-field strength is large enough to influence the mode volume of our system. While the Q-factor still decreases in this case, the simultaneous decrease of $V_{mod}$ (down to $V_{mod} = 3.71(\lambda/n)^3$ for $r = 35$ nm) overcompensates this effect. For diamond spheres positioned (3) in the evanescent field next to the waveguide, we find qualitatively similar behavior as in case (1) and for placement (4) on top of the $Ta_2O_5$ waveguide $Q/V_{mod}$ remains constant for all $r$. Generally, we find $Q$ to be significantly more sensitive to changes in the geometry as compared to $V_{mod}$. Hence, $Q$ can decrease as a result of the perturbation, while the relative E-field strength at the position of the sphere in these cases is too low to affect $V_{mod}$.

In none of the cases do we observe any significant shift of the resonance wavelength $\lambda_{res}$ ($\Delta\lambda_{res} < 0.1$ nm). Based on this numerical analysis, position (2) will be favorable in experimental realizations. (Note however that $Q/V_{mod}$ would decrease in cases where $r$ increases beyond the maximum at $r = 35$ nm).

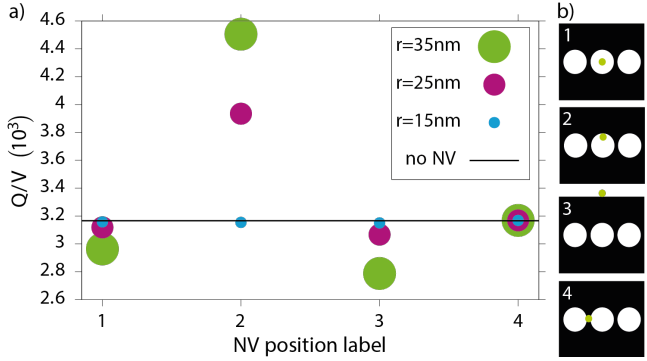

Figure 3: a) $Q/V_{mod}$ ratio of the PhC cavity perturbated by a nanodiamond sphere (radius $r_{dia} = 15, 25, 35$ nm). Unperturbated cavity $Q/V_{mod}$ ratio for reference. The diamond sphere is placed in four different positions close to the defect center labeled 1-4. b) Schematic illustration of the dielectric emitter-cavity system for case 1-4.

## III. FABRICATION

Based on the above parameter optimization, we fabricate devices in multiple layers of electron beam lithography (EBL). The nanophotonic platform is developed from commercial silicon wafers with a 2.6 µm thick thermal silicon dioxide layer onto which a 330 nm thin layer of $Ta_2O_5$ was deposited in a sputtering process. (i) Gold markers for the alignment of subsequent lithography layers with sub 50 nm overlay accuracy are fabricated in a lift-off process. (ii) Grating couplers, waveguides and the PhC cavities are patterned in high resolution electron-beam resist (man 2403) on a 100 kV Raith EBPG 5150 system. After resist reflow, reactive ion etching transfers the pattern into the $Ta_2O_5$ thin film.[23] (iii) We then spin-coat the chip with a thin film of PMMA and use EBL to pattern apertures of 240 nm diameter at the center hole of each PhC cavity. Nanodiamonds in a DI water solution (micro-diamant, unprocessed) are then drop-casted onto the PMMA-aperture pattern and we perform a lift-off process in acetone to remove excess emitters. Whereas on the order of 20 nanodiamonds typically remain within the aperture at the center PhC cavity hole, we only find one of them hosting an active and stable NV center on average. (vi) Finally, we scan over the device multiple times at high excitation powers (>10 mW) using a confocal microscope in order to bleach the waveguide material, reduce auto-fluorescence from $Ta_2O_5$ and eliminate unstable NV centers that can occur close to the surface of a nanodiamond.

## IV. PURCELL ENHANCEMENT

Fabricated devices are characterized in confocal microscopy in order to assess, if the nanodiamonds at the center of the cavity contain photo-active NV centers. For this purpose, we record the photoluminescence under cw-illumination with a high power (>10 mW) off-resonant (532 nm excitation wavelength) laser through a 100x objective with NA = 0.9. A dichroic mirror (Semrock, 595 nm) allows for separating excitation and photoluminescence light and 650 nm long-pass and 750 nm short-pass optical filters select the spectral region of interest. The spectral and statistical properties of the collected light field are then analyzed in a single photon sensitive spectrometer (Andor Shamrock with iDus CCD camera) and a Hanbury-Brown-Twiss (HBT) setup, respectively.

For illumination of the devices with high optical power we observe that the self-fluorescence from the $Ta_2O_5$ waveguides bleach and saturate at levels approximately five times inferior to the count rate we record from a single NV center, as shown in fig. 1 b). Recording the photoluminescence when scanning the device in the lateral directions allows us to confirm optimal positioning of NV centers with respect to the center of the PhC cavity and an additional waveguide approaching perpendicularly in a cross-bar geometry as shown in figure 1 a-b).

Next, we study the Purcell effect for NV centers embedded into 1D PhC cavities and compare the emission rates in the cavity-coupled (resonant) $\gamma_{1,c}$ and uncoupled, reference case, $\gamma_{1,ref}$, which gives rise to the total Purcell enhancement factor $F_{tot} = \frac{\gamma_{1,c}}{\gamma_{1,ref}}$. The corresponding modifications of the brightness, the spectrum and the lifetime of the NV center allow for independent determinations of the Purcell enhancement, as discussed in the following.

An intuitive approach to determining the Purcell enhancement relies on comparing the emission rates of cavity-coupled and uncoupled emitters. The measured photoluminescence intensity $I$ of an emitter approaches the saturation intensity $I_{sat}$ at the saturation power $P_{sat}$, following a saturation behavior[13]:

$$I(P) = \frac{I_{sat}}{\left(1 + \frac{P_{sat}}{P}\right)} \quad (1)$$

Note that this equation assumes a quantum efficiency of unity, which will not generally be achieved at elevated excitation powers. The saturation intensity here depends on the collection efficiency and therefore on the directionality of the emission. In our case the 1D PhC enhances the emission into the waveguide, rather than into the microscope objective, such that the measured intensity is not directly informative of the Purcell enhancement. However, the saturation power is related to the radiative lifetime and therewith allows for determining the total enhancement factor from comparing cavity-coupled and reference values under similar excitation conditions (see Appenix),

$$F_{\text{tot}} = \frac{P_{\text{sat,c}}}{P_{\text{sat,ref}}} \quad (2)$$

An increase of the radiative decay rate allows for faster repopulation of the excited state, which then results in an increased saturation power.

We experimentally determine the saturation behavior of individual NV centers by performing photoluminescence scans at different excitation powers. In fig. 4 a) we show the result of multiple such scans for an NV center on a far-detuned cavity, thus representing the uncoupled case for reference. Each data point is a weighted average of multiple scans, for which we consider 10% of the lowest intensity data as background and fit the signal with a two-dimensional Gaussian fit.

The uncoupled NV center on a far-detuned PhC cavity serves as our reference device and we find $I_{\text{sat,ref}} = 106 \pm 10$ kcnt/s and $P_{\text{sat,ref}} = 3.6 \pm 0.7$ mW from a fit to the data shown in fig. 4 b), which is similar to reported values in the literature[24]. For the cavity-coupled device, we find that the saturation intensity is on the same order of magnitude with $I_{\text{sat,c}} = 140 \pm 28$ kcnt/s. The saturation power, however, increased to $P_{\text{sat,c}} = 11.0 \pm 3.8$ mW, which results in a total enhancement factor of $F_{\text{tot,sat}} = 3.1 \pm 1.1$. Note that we here neglect the non-radiative decay path and assume that both NV centers have similar properties, although they are located in different nanodiamonds.

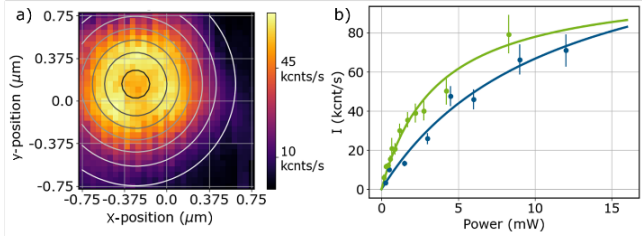

Figure 4: Saturation behavior of photoluminescence from NV centers a) Average photoluminescence signal recorded in confocal microscopy by scanning the sample in lateral x- and y directions. b) photoluminescence signal vs. excitation power of uncoupled NV center (green, $I_{\text{sat,ref}} = 106 \pm 10$ kcnt and $P_{\text{sat,ref}} = 3.6 \pm 0.7$ mW) and cavity-coupled NV center (blue: $I_{\text{sat,c}} = 140 \pm 28$ kcnt/s and $P_{\text{sat,c}} = 11.0 \pm 3.8$ mW).

We further extract the total enhancement factor from comparing the spectral power distribution ($SPD(\lambda_{\text{res}})$) of photons emitted from the cavity-coupled system with those emitted from an uncoupled NV center. We here assume that the emission rate will only be enhanced for transitions resonant with the PhC cavity and normalize the spectral power distribution of the cavity-coupled device to compensate for the reduced collection efficiency, $\zeta_c$, in confocal microscopy. We can then write the total enhancement factor as[5,25]

$$F_{\text{tot}} = \frac{\zeta_{\text{ref}}}{\zeta_c}\left(\frac{SPD_c(\lambda_{\text{res}})}{SPD_{\text{ref}}(\lambda_{\text{res}})}\right) \approx \frac{SPD_c(\lambda_{\text{res}})}{SPD_{\text{ref}}(\lambda_{\text{res}})} \quad (3)$$

We record spectra from an ensemble of uncoupled NV centers and the single PhC cavity-coupled NV center and find that the emission of the latter is dominated by a sharp resonance close to the design wavelength of the cavity at 705 nm (see fig. 5 a). The SPD is then obtained from subtracting the dark counts, dividing the resulting data by the sum of all counts and normalizing to the recorded bandwidth. We find that the SPD of the emission from reference NV centers increases from $SPD_{\text{ref}} = 1.05 \pm 0.05$ %/nm to $SPD_c = 2.20 \pm 0.05$ %/nm for the PhC cavity coupled device. Note however, that the recorded spectrum is limited by the resolution of our spectrometer ($\sigma = 0.55$ nm). The corresponding broadening of the observed cavity resonance can be taken into account through fitting the data with a Voigt profile but requires knowledge of the cavity quality factor. We determine the latter by using a higher resolution spectrometer (of limited spectral range) and find $Q = 3799 \pm 230$ from a fit to the data shown in fig. 5 b). Fitting the data of the resonant feature in the broader bandwidth spectrum with the Voigt profile, as shown in fig. 5 c), then yields $SPD_c = 3.75 \pm 0.14$ %/nm, resulting in an enhancement factor of $F_{\text{tot}} \approx 3.6 \pm 0.2$.

While this independently obtained result is consistent with our previous determination of the Purcell enhancement via the saturation behavior and the analysis of spectral power densities is well-established, we note that the method relies on the assumption that the ensemble of NV centers used as a reference has similar spectral properties as the PhC cavity coupled NV center. However, the photophysical properties of nanodiamonds strongly depend on their local (strain and electrical) environment, which may cast doubt on the validity of the assumption. We hence sought yet another independent assessment of the Purcell enhancement via the temporal properties of the emitted photoluminescence.

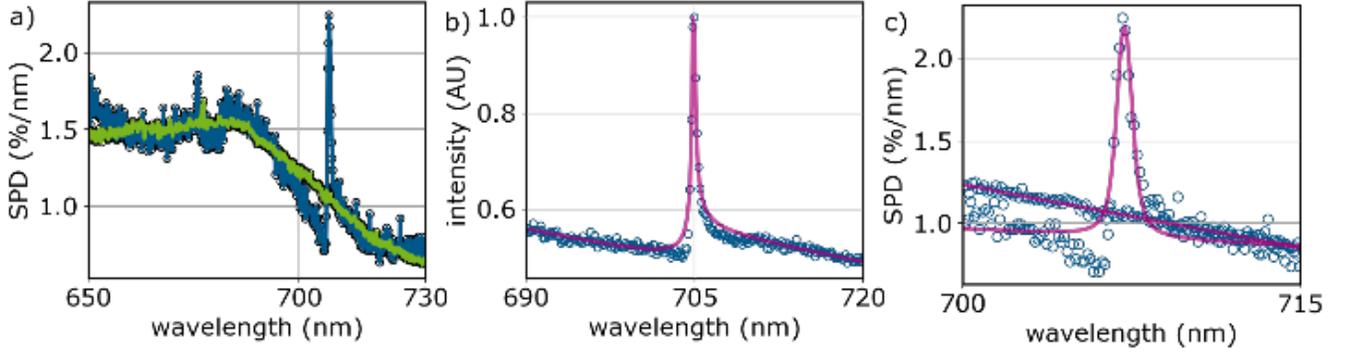

Figure 5: Spectral photoluminescence measurements a) spectral power distribution in percent of the cavity-coupled device (blue) and reference ensemble of NV centers (green) normalized to total number of counts and bandwidth. At the resonance wavelength (of 705 nm) the spectral power distribution is 1.05 %/nm for un-coupled NV centers while it reaches 2.2 %/nm for the cavity-coupled device, which is limited by the resolution of the spectrometer. b) Lorentz fit to the data for the cavity-coupled device on resonance, recorded with a high resolution grating spectrometer, yielding Q= 3799 +-230. c) Voigt profile fit to the data recorded with a low-resolution spectrometer yields a peak intensity of $SPD_c = 3.75 \pm 0.14\ \%/nm$ .

The most direct way to measure the Purcell enhancement is by assessing the enhancement of the radiative lifetime $\tau_1 = 1/\gamma_1$. For unity quantum efficiency the total enhancement factor then reads (see Appendix)

$$F_{tot} = \frac{\tau_{1,ref}}{\tau_{1,c}} \quad (4)$$

Whereas the lifetime in bulk diamond consistently yields values of 12 ns,[2] in nanodiamonds effects arising from the local electrical and strain environment must be considered and complicate the determination of $\tau_{1,ref}$. Common methods for determining the lifetimes of uncoupled emitters in nanodiamonds include actively detuning the cavity resonance and physically displacing the emitter from the cavity, which however also leads to changes in the dielectric environment among other variations that cause lifetime modifications.

It is not straightforward to tune integrated PhC cavities over a wide frequency range and physical displacements of nanodiamonds on a chip are extremely cumbersome. For broadband emitters like the NV center, we instead perform an in-situ characterization by spectral selection. For this purpose, we separate the spectrum into an interval dominated by resonant cavity features (650-750 nm) and a reference interval far-off resonance (630-640 nm). We then perform measurements of the second order auto-correlation function ($g^{(2)}(\tau)$) using the HBT-setup and record the photon-photon waiting time distribution with time delay $\tau$ between single photon detection events and apply a three level model (see fig. 6 a) to fit the data.

The measurement of the device in the cavity dominated interval shows a minimal value at vanishing time-delay ($\tau = 0$) of $g^{(2)}(0) \approx 0.3$ without background correction, as shown in fig. 6 b). Here values below $g^{(2)}(0) < 0.5$ validating that the recorded light originates from a single NV center. Deviations from the ideal case of $g^{(2)}(0) = 0$ are due to the limited timing jitter of the single-photon detectors (~350 ps) and the presence of detection events from incoherent background photons. The optical nutation beyond values of $g^{(2)}(\tau) = 1$ in the vicinity of $\tau = 0$ can be explained by the presence of a non-radiative third level or shelving state (fig. 6 a).

A suitable functional description of the antibunching behavior is hence based on a three-level model, which we correct for uncorrelated background, and reads[13]

$$g^{(2)}(\tau) = 1 - \beta e^{-\gamma_1 \tau} + (1-\beta)e^{-\gamma_2 \tau} \quad (5)$$

The first term of the above equation describes the anti-bunching of photons emitted within a two-level sub-system of $\tau_1 = 1/\gamma_1$ lifetime. The second term allows for reproducing bunching behavior originating from a non-radiative third level (shelving state) with the lifetime $\tau_2 = 1/\gamma_2$. The coefficient $\beta$ indicates the contribution of such shelving state to the overall behavior. We take into account uncorrelated background contributions by modulating the above $g^{(2)}$-function with a parameter $\rho = \frac{\text{signal}}{\text{signal+noise}} = \frac{\text{counts}}{\text{counts+background}}$, where signal and background are measured as reported for the saturation curve. Employing such a corrected correlation function $g^{(2)}_{corr}(\tau) = 1 + \frac{(g^{(2)}(\tau)-1)}{\rho^2}$ a fit to the raw data yields $g^{(2)}_{corr}(0) = 0.20 \pm 0.16$, which now takes into account background contributions. From the fit we further extract the lifetime of the coupled system $\tau_{coupled} = 1/\gamma_c = 4.1 \pm 0.4$ ns.

For the reference measurement far-off resonance not only a lower count rate due to the significantly smaller spectral bandwidth (630-640 nm) is expected, but also the background contribution of the $Ta_2O_5$ material is considerably higher, as evident from fig. 6 c). In order to separate the background from the contributions of a single NV center to the overall count rate, we include the $\rho$ factor as a free fit parameter and enforce $g_{corr}^{(2)}(0) = 0$ as a condition for a background-free ideal quantum system. The resulting fit matches the data well and reveals a lifetime of $\tau_{ref} = 1/\gamma_{ref} = 11.1 \pm 2.8$. Comparing the cavity-coupled and uncoupled (reference) cases, we find a Purcell enhancement of $F_{tot,\tau} = 2.7 \pm 0.5$.

In summary, the determination of the total enhancement factor with three independent methods consistently shows a Purcell enhancement significantly above the vacuum contribution. The determination via the lifetime yielded a slightly lower value for the total enhancement factor as compared to the other two methods because here the internal quantum efficiency was not considered, which we study in further detail via investigating the dynamics of the coupled system.

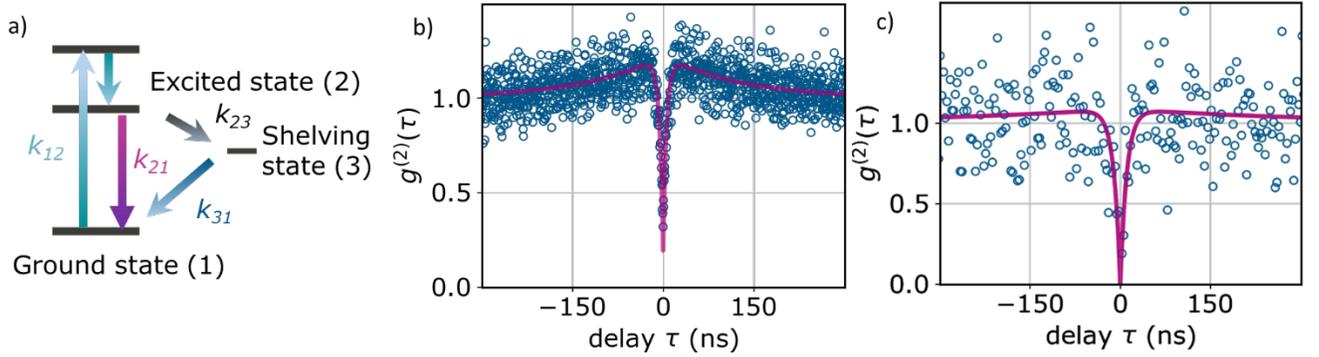

Figure 6: Auto-correlation $g^{(2)}(\tau)$ measurements and three level model. a) Three level model of the NV center. The system is optically excited by a 532 nm wavelength laser from the ground state (1) to the excited state (2) with the rate $k_{12}$. The system decays with radiative rate $k_{21}$ to the ground state and with non-radiative rate $k_{23}$ to the shelving state (3). The de-shelving rate $k_{31}$ resets the system back to the ground state. b) Auto-correlation measurement in cavity-coupled interval with spectral filtering between 650-750 nm. (raw data in blue and 3-level fit in purple) with lifetime of $\tau_c = 4.1 \pm 0.4$ ns. c) Auto-correlation measurement in reference interval with spectral filtering in the 630-640 nm range. (background corrected data in blue and 3-level fit in purple) with lifetime of $\tau_{ref} = 11.1 \pm 2.8$ ns. Both measurements are performed with a 532 nm wavelength (3 mW) excitation laser.

## V. QUANTUM EFFICIENCY

We consider the emitter as a three-level system with a radiative and a non-radiative decay path in order to specify the internal quantum efficiency. For simplicity all non-radiative transition rates are summarized in a single rate into a shelving state,[13] as shown in Figure 6 c). In this model the excitation from the ground (1) to the excited state (2) occurs with the rate $k_{12}$, which depends on the laser power. Photon emission from the excited to the ground state, correspondingly, occurs at rate $k_{21}$ and all non-radiative decays into the shelving state (3) are summarized by rate $k_{23}$. The de-shelving rate $k_{31}$ expresses reinitialization of the system into the ground state. We hence define the quantum efficiency as the ratio of the radiative to the total (radiative and non-radiative) rate:

$$Q_{int} = \frac{k_{21}}{k_{21} + k_{23}} \quad (6)$$

While the transition rates cannot be measured directly, it is possible to deduce them from measurements of the second order auto-correlation function. We here express the rate of detected photons $R$, the inverse lifetimes $\gamma_1, \gamma_2$ and the shelving state contribution $\beta$, as determined from the fitted $g^{(2)}(\tau)$ functions, in terms of the four transition rates $k_{ij}$. For the case of small non-radiative contributions, we find [13]:

$$R \approx \zeta k_{21} \left(\frac{k_{12}}{k_{12} + k_{21}}\right)$$
$$\gamma_1 \approx k_{21} + k_{12}$$
$$\gamma_2 \approx k_{13} + k_{23}\left(\frac{k_{12}}{k_{12} + k_{21}}\right)$$
$$\beta \approx 1 + \frac{k_{23}}{k_{31}}\left(\frac{k_{12}}{k_{12} + k_{21}}\right) \quad (7)$$

Determination of the setup efficiency $\zeta$ requires a careful calibration of the setup, which we here achieve by exploiting the linear correlation between excitation rate and laser power ($k_{12} \sim P$), as described previously by Berthel et al.[13]

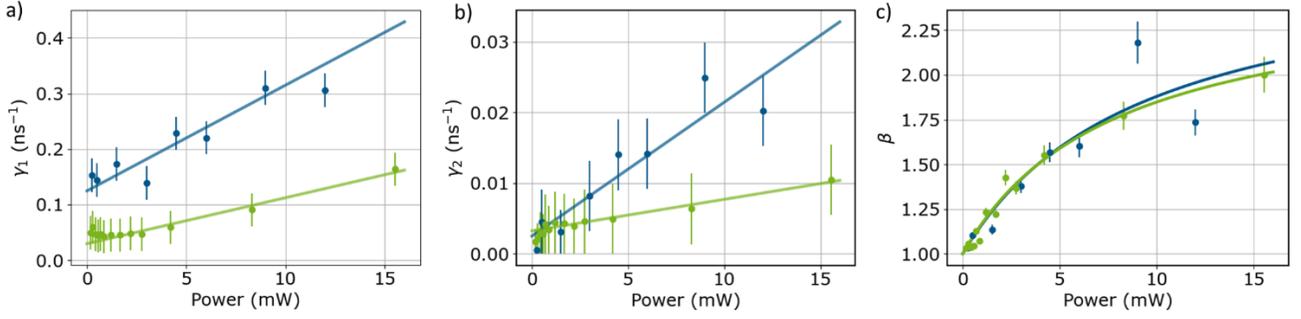

Figure 7: Parameters from 3-level-fit for an uncoupled (green) and a cavity-coupled device (blue) for different excitation powers: a) decay rate of anti-bunching term $\gamma_1$ with linear fits to the data, b) decay rate of the bunching term $\gamma_2$ with linear fits to the data and c) coefficient of the shelving state contribution $\beta$ with a saturation fit model $\beta(P) \approx 1 + a\left(\frac{1}{1+b/P}\right)$.

We measure second order auto-correlation functions with excitation powers of up to 15 mW for the same devices for which we determined the power dependence of the detection rate shown in fig. 4 b) and extract the coefficients $\gamma_1, \gamma_2$ and $\beta$, taking into account all three levels as considered in fig. 6 a). In order to provide realistic uncertainties, we account for the uncertainty of the fitting routines, the binning of the measurement data and the signal to noise ratio.

The corresponding data for the cavity-coupled and the reference NV center extracted from the measurements of the $g^{(2)}$-function is shown in fig. 7 in dependence of the excitation power. In addition to the previous mentioned enhancement of the radiative rate for the cavity-coupled device, we observe similar power dependences of the fit parameters for both devices. Notably, the $\gamma_1$ and $\gamma_2$ coefficients follow a linear dependence at higher excitation powers, whereas the $\beta$ coefficient shows a saturation behavior.

The observed power dependencies of the coefficient already allow us to estimate the power dependencies of the rates $k_{ij}$. Under the assumptions that (i) the radiative decay rate is independent of the excitation power $k_{21}(P) \approx const.$ and that (ii) the power dependence of the excitation rate is linear $k_{12} \sim P$, we can explain the linear increase of $\gamma_1(P) \approx k_{21} + k_{12}(P)$. The linear power dependence of $\gamma_2(P)$ implies a dominant power dependence of the non-radiative rates $k_{23}(P)$ or $k_{31}(P)$, because the $k_{12}(P)$-term is expected to show saturation behavior [see Eq. (7)]. The saturation behavior of the $\beta$ coefficient further suggests, that the power dependence of $k_{23}(P)$ and $k_{31}(P)$ is of the same order, so that the curve is governed by the saturation of the $k_{12}(P)$ term. Whereas all fit functions in fig. 7 agree reasonably well with the data, the linear regression of $\gamma_1(P)$ requires the least stringent assumptions and was used to calibrate the setup efficiency $\zeta$. For the uncoupled NV center we find a reference value of $\zeta_{ref} = 15 \cdot 10^{-3}$ and for the cavity-coupled device we obtain $\zeta_c = 2 \cdot 10^{-3}$ (see Appendix for details).

Using the setup efficiencies, the detected photoluminescence rates (fig. 4) and the coefficients describing the antibunching behavior (fig. 7), we can now calculate the coefficients $k_{ij}$, as shown in figure 8. Note that we use the exact formulas provided in [13], which remain valid for higher excitation powers, rather than Eq. (7) for these final calculations and added an uncertainty of 10% for the determined efficiency $\zeta$.

The calibration procedure allows us to access the intrinsic $k_{21}$ radiative decay rate, which is otherwise obscured by the power dependence of the measured parameters. In fig. 8 a) we see that a linear fit to the data for the cavity-coupled device (blue) approaches the weighted mean of the data points below the saturation power (dotted). In the case of the uncoupled device (green), we find that the data points are scattered around the weighted mean at low excitation powers, but a linear power dependence is observed above the saturation power (as also reported elsewhere[13]). In order to determine the Purcell enhancement when including non-radiative transitions and larger $k_{12}$ contributions, we use the weighted average of the $k_{21}$ values below saturation power, yielding $k_{21,ref}(P_0) = 0.039 \pm 0.005$ ns$^{-1}$ = $1/(25 \pm 3)$ ns and $k_{21,c}(P_0) = 0.140 \pm 0.019$ ns$^{-1}$ = $1/(7.1 \pm 1.0)$ ns, and find $F_{tot,k} = \frac{k_{21,c}(P_0)}{k_{21,ref}(P_0)} = 3.6 \pm 0.5$, which agrees with our previous observations.

The excitation rate $k_{12}$ increases linearly for both cavity-coupled and uncoupled devices, as seen in fig. 8 b), in accordance with our previous considerations. The rates for the uncoupled device remain below previous reported values, which we attribute to differences in the numerical aperture of the optical systems[26]. The cavity-coupled device shows an enhanced excitation rate, which is consistent with a Purcell enhancement of the radiative rate.

The non-radiative rates $k_{23}$ are negligible for experiments with low excitation power in both cavity-coupled and uncoupled cases but increase at higher excitation powers, as shown in fig. 8 c). We find that the rates are similar for both devices ($k_{23,c} \approx k_{23,\text{ref}}$), which leads us to the conclusion that the dielectric environment of the cavity as well as the Purcell effect have no significant effect on the non-radiative decay paths. Moreover, both data sets follow an approximately linear trend in accordance with previous considerations. However, the non-radiative rate remains smaller than the radiative rate $k_{23} < k_{21}$ for all considered excitation power levels.

We further observe the de-shelving rate $k_{31}$, shown in fig. 8 d), which seems to show saturation behavior at relatively low excitation power for the uncoupled NV center (possibly limited by the excitation rate $k_{12} \geq k_{31}$), while the cavity-coupled device keeps monotonically increasing (possibly due to increased saturation power by the Purcell effect). More detailed studies will be required to elucidate the de-shelving behavior in both cases.

In summary, the determination of the rates generally agrees with our model conceptions, as underlined by the total enhancement factor, found from comparing the radiative decay rates and will be used in the following for calculating the internal quantum efficiencies in dependence of excitation power.

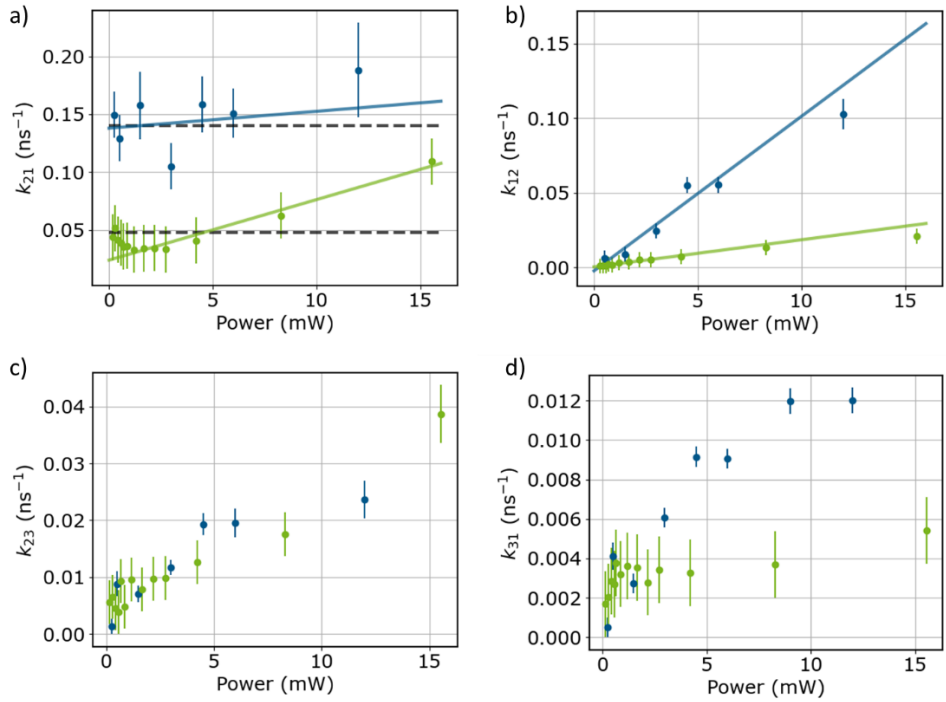

Figure 8: Three-level rates for uncoupled reference (green) and cavity-coupled NV center (blue). a) The radiative transition rate $k_{21}$ is increased by the total enhancement factor for the cavity-coupled device, which is evident from the weighted average for the data below the saturation power as black dotted line (3.6 mW). The power dependence is reduced in comparison to Fig. 7a), but an increase at elevated powers is still visible as visualized by the linear fit to the data above the saturation power (solid line) b) The excitation rate $k_{12}$ follows a linear power dependence over the whole range, as shown by the fit (solid line). c) The non-radiative rate $k_{23}$ is similar for both devices and independent from cavity coupling (Purcell effect). d) The de-shelving rate $k_{31}$ seems to saturate for the uncoupled reference device but increases monotonically for the cavity-coupled device.

With the radiative and non-radiative rates determined from fig. 8 a) and c), we derive the internal quantum efficiency [Eq. 6] for the different excitation powers, as shown in fig. 9. We find that the efficiency of the uncoupled NV center saturates at $Q_{\text{int,ref}} \approx 75\,\%$ for excitation powers approaching saturation, which agrees with values of around 70 % for NV- centers in nanodiamonds reported elsewhere[12]. In contrast to this, the internal quantum efficiency of the PhC cavity-coupled NV center remains at 90 % even for high excitation powers (see fig. 9), as a consequence of the Purcell effect.

While the radiative decay rate is enhanced by the cavity $k_{21,c} \approx F_{\text{tot}} k_{21,\text{ref}}$ (as discussed in the previous section), the non-radiative decay rate remains unaffected $k_{23,c} \approx k_{23,\text{ref}}$. When we apply these findings in the definition of $Q_{\text{int}}$, we can relate the quantum efficiencies of cavity-coupled and uncoupled devices via the total enhancement factor[27]:

$$Q_{\text{int,c}} \approx \frac{1}{1+\frac{k_{23,c}}{k_{21,c}}} = \frac{1}{1+\frac{1}{F_{\text{tot}}}\frac{k_{23,\text{ref}}}{k_{21,\text{ref}}}}$$
$$= \frac{1}{1+\frac{1}{F_{\text{tot}}}\left(\frac{1-Q_{\text{int,ref}}}{Q_{\text{int,ref}}}\right)} \quad (8)$$

which for high internal quantum efficiencies $Q_{\text{int}} \to 1$ approaches

$$Q_{\text{int,c}} \approx 1 - \left(\frac{1}{F_{\text{tot}}}\right)(1-Q_{\text{int,ref}})$$
or
$$\frac{1-Q_{\text{int,ref}}}{1-Q_{\text{int,c}}} \approx F_{\text{tot}}$$

Hence, the Purcell enhancement directly expresses deviations from ideal internal quantum efficiencies between both cases.

If we apply this finding to our measurement, where the uncoupled NV center in a nanodiamond has a quantum efficiency of $Q_{\text{int,ref}} \approx 75\%$, the coupling to the PhC cavity with $F_{\text{tot}} \approx 3$ would result in an increased internal quantum efficiency of $Q_{\text{int,c}} \approx 92\%$.

We can further employ the quantum efficiency determination for improving the previously discussed calculation of the Purcell factor via data on the radiative lifetimes. Here we find that

$$\frac{\tau_{\text{ref}}}{\tau_c} = \frac{k_{21,c}+k_{23,c}}{k_{21,\text{ref}}+k_{23,\text{ref}}} = \frac{\frac{k_{21,c}}{k_{21,\text{ref}}}+\frac{k_{23,\text{ref}}}{k_{21,\text{ref}}}}{1+\frac{k_{23,\text{ref}}}{k_{21,\text{ref}}}}$$
$$= \frac{F_{\text{tot},\tau,\text{cor}}+\left(\frac{1-Q_{\text{int,ref}}}{Q_{\text{int,ref}}}\right)}{\frac{1}{Q_{\text{int,ref}}}}$$

and correspondingly the quantum efficiency corrected Purcell factor,

$$F_{\text{tot},\tau,\text{cor}} = \frac{1}{Q_{\text{int,ref}}}\left(\frac{\tau_{\text{ref}}}{\tau_c} - (1-Q_{\text{int,ref}})\right)$$

With the previously determined ratio of $\frac{\tau_{\text{ref}}}{\tau_c} = 2.7 \pm 0.5$ and the internal quantum efficiency of $Q_{\text{int,ref}} = 0.75$, we find a corrected total enhancement factor of $F_{\text{tot},\tau,\text{cor}} = 3.3 \pm 0.6$.

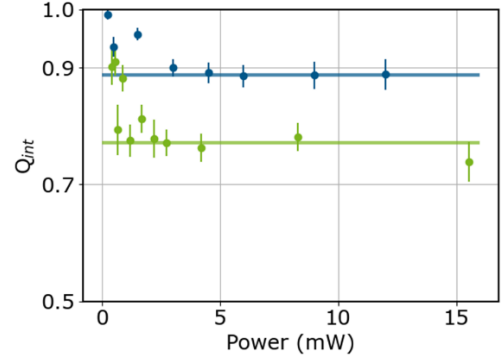

Figure 9: Quantum efficiency over excitation power for the uncoupled (green) and the cavity-coupled NV center (blue). Solid lines show the weighted average of the measurements above the saturation power of the uncoupled device (3.6 mW), where the quantum efficiencies are saturated for both devices.

## VI. CONCLUSION

In summary, we have studied the Purcell effect on the internal quantum efficiency for an NV center in a nanodiamond coupled to a 1D PhC cavity. Photoluminescence and second order auto-correlation measurements on single NV centers in nanodiamonds positioned at the center of the cavity allowed us to determine the Purcell enhancement via three independent methods. Assessment of the saturation behavior, the spectral power density and the lifetime yielded $F_{\text{tot}} = (3.1 \pm 1.1, 3.6 \pm 0.2, 3.3 \pm 0.6)$, respectively. All methods confirm a significant coupling between emitter and cavity. Measurements of the auto-correlation function at elevated excitation power levels further allowed us to extract the efficiency of the optical setup as well as the radiative and non-radiative decay rates of our NV center system within a three-level model description. We found experimentally that whereas the radiative decay rate is increased by the total enhancement factor, the non-radiative decay rate remains unaffected by the cavity coupling. Importantly, our studies benefitted from low loss transmission and low self-fluorescence of the employed $Ta_2O_5$-on-insulator nanophotonic platform, in contrast to loss-dominated studies with plasmonic structures.[9,16]

By combining our findings, we discover that the Purcell effect enhances the internal quantum efficiency from $Q_{\text{int,ref}} \approx 0.75$ for an uncoupled emitter to $Q_{\text{int,c}} \approx 0.9$ for a single cavity-coupled NV center, in accordance with related work on SiV centers.[27]

Engineering the optical properties of waveguide-integrated single photon emitters is an important aspect of scalable quantum optics experiments. Our results elucidate the complex photophysical properties of NV-centers in nanodiamonds that couple to nanophotonic resonators and will aid the design of efficient integrated single photon sources with high internal quantum efficiencies and high repetition rates, as desired for many quantum technology applications, such as quantum key distribution.

## ACKNOWLEDGEMENTS

We thank the Münster Nanofabrication Facility (MNF) for support in fabrication matters and Dr. Johannes Kern for fruitful discussions. C.S. acknowledges support from the Ministry for Culture and Science of North Rhine-Westphalia (421-8.03.03.02–130428).

# APPENDIX

## 1. Purcell factor from measurements

**Saturation behavior:**

The steady state solution for the excited state of a two level quantum emitter can be written as [9]:

$$\rho_2 = \frac{k_{12}}{k_{12} + k_{21}}$$

Where $\rho_2$ is the normalized state population of the excited state, $k_{12}$ the excitation rate and $k_{21}$ the radiative emission rate. Because the excitation rate is linearly dependent on the excitation power ($k_{12} = \sigma_{abs}P$) and the detected photoluminescence intensity $I(P)$ is proportional to $\rho_2$, we can rewrite the above expression as[9]:

$$I(P) = I_0 \frac{1}{1 + \frac{k_{21}}{\sigma_{abs}P}}$$

Following we identify the saturation power of the initial equation as $P_{sat} = \frac{k_{21}}{\sigma_{abs}}$. Combined with the definition of the total enhancement factor we find [28]:

$$F_{tot} = \frac{k_{21,\,coup}}{k_{21,\,uncoup}} = \frac{P_{sat,\,coup}}{P_{sat,\,uncoup}}$$

Note that for this derivation all non-radiative decay channels are neglected[29] and the absorption cross-section ($\sigma_{abs}$) is assumed to be equal for different emitters.

**Voigt fit:**

The amplitude $A(PSB_{res})$ of the resonance is restored with a stretching factor of $\frac{1}{\frac{\sigma}{2.3548}\sqrt{2\pi}} = 1.88$, such that we find that the $PSB_{coupled} = \frac{1}{\frac{\sigma}{2.3548}\sqrt{2\pi}} A(PSB_{res}) + BG(PSB_{res}) \approx 1.88 \cdot PSB_{res} + 1$.

**Purcell theory:**

In order to estimate the Purcell enhancement, we measure the spectrum with a high resolution grating and a small aperture and fit the resonance by a Lorentzian function, which yields a Q-factor of $Q = 3799 \pm 133$. The Purcell factor is defined as the ratio of the decay rate into the cavity $\gamma_c$, in reference to the free space decay rate $\gamma$:

$$F_{P,theo} = \frac{\gamma_c}{\gamma} = \frac{3}{4\pi^2}\left(\frac{\lambda_c}{n}\right)^3 \frac{Q}{V_{mod}} \zeta_{imp}$$

Where $\lambda_c/n$ is the resonance wavelength of the cavity in the waveguide material, $Q$ is the quality factor and $V_{mod}$ is the mode volume of the cavity. The additional factor $\zeta_{imp}$ incorporates experimental conditions like the detuning and spatial overlap between cavity and emitter, thus reducing the maximal achievable Purcell enhancement. Under ideal conditions ($\zeta_{imp} = 1$), the measured Q-factor and the simulated mode volume of $V_{mod} = 3.71 \left(\frac{\lambda_c}{n}\right)^3$, we calculate $F_{P,max} \approx 78$.

If we further consider the detuning of the cavity to the 4th PSB, the value is reduced by a factor of $\zeta_{det} \sim 0.75$ to $F_{P,det} = 58$.

For a negatively charged NV center at room temperature the emission is governed by dephasing ($\gamma^* \approx$ THz), which puts the system in the "bad emitter"-regime, where the linewidth of the cavity cancels out and the above values only serve as an upper limit and prospect for low temperature measurements.

Whereas the theoretical Purcell factor is only defined for a single wavelength and coupling to a single sideband, we define the total enhancement factor $F_{tot}$ for the measured spectrum.

$$F_{tot} = \sum \frac{\gamma_{c,i}}{\gamma} = \frac{\gamma_{c,tot}}{\gamma}$$

If the coupling to a single transition band with a branching $\zeta_n \gg \zeta_i$ is dominant it can be related to the Purcell factor as:

$$F_{tot} \approx \frac{\gamma_{c,n}\zeta_n + \gamma}{\gamma} = F_P\zeta_n + 1$$

## 2. Determination of setup efficiency: (Fig 7 a-d)

The decay rates of the three levels can be deduced by following the procedure proposed by Berthel et al. The rates are calculated from the derived parameters $\gamma_1$, $\gamma_2$ and $\beta$ and the recorded count rate R combined with the setup efficiency $\zeta$. The setup efficiency is initially unknown and different for every measurement (random dipole orientation) but is determined by varying $\zeta$ around an estimated value while calculating the $k_{21}$-slope and -y-intercept.

We find for the reference measurement of the uncoupled device that the $k_{21}$ y-intercept approaches the one of $\gamma_1$ with a minimal $k_{21}$ slope for a setup efficiency of $\zeta_{ref} = 15 \cdot 10^{-3}$ (Figure A1 a), which is larger than for Berthel et al. because of the larger filter window and different optics. For the cavity-coupled device the y-intercept is best satisfied for $\zeta_c = 2 \cdot 10^{-3}$ (Figure A1 b), which is almost an order of magnitude lower than the uncoupled reference device. This is a further indication that most of the light is directed into the cavity. The parameters are then used to determine the rates of the 3-level system.

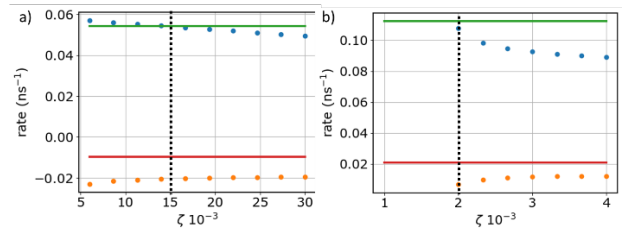

Figure A1: Modification of setup efficiency $\zeta$. a) $\gamma_1$ fit parameters from the reference NV with- y-intercept (green) and -slope (red) from linear fits and $k_{21}$- y-intercept (blue) and -slope (orange) from linear fits. b) same as a) for the coupled device.